\documentclass[twoside,11pt]{article}

\usepackage{jmlr2e}
\usepackage{framed,multirow}
\usepackage{amssymb}
\usepackage{latexsym}
\usepackage{amsmath}
\usepackage{booktabs}
\usepackage{array,multirow,graphicx}
\usepackage{stackengine}
\usepackage{relsize}
\usepackage[inline]{enumitem}
\usepackage[ruled,vlined]{algorithm2e}
\usepackage{graphicx}
\usepackage[skip=2pt]{subcaption}
\usepackage{amssymb}
\usepackage{pifont}
\usepackage{mwe}
\usepackage{paralist}
\usepackage{xcolor}
\usepackage{soul}

\usepackage[acronym]{glossaries}
\makeglossaries
\newacronym{wsi}{WSI}{Whole Slide Image}
\newacronym{he}{H\&E}{Hematoxylin and Eosin}
\newacronym{cnn}{CNN}{Convolutional Neural Network}
\newacronym[plural=GNNs,firstplural=Graph Neural Networks (GNNs)]{gnn}{GNN}{Graph Neural Network}
\newacronym{mil}{MIL}{Multiple Instance Learning}
\newacronym{knn}{k-NN}{k-Nearest Neighbors}
\newacronym[plural=RAGs,firstplural=Region Adjacency Graphs (RAGs)]{rag}{RAG}{Region Adjacency Graph}
\newacronym[plural=RoIs,firstplural=Regions of Interest (RoIs)]{roi}{RoI}{Region Of Interest}
\newacronym[plural=TRoIs,firstplural=Tumor RoIs (TRoIs)]{troi}{TRoI}{Tumor RoI}
\newacronym{ui}{UI}{User Interface}
\newacronym{api}{API}{Application Programming Interface}
\newacronym{ml}{ML}{Machine Learning}
\newacronym{dl}{DL}{Deep Learning}
\newacronym{ai}{AI}{Artificial Intelligence}
\newacronym{slic}{SLIC}{Simple Linear Iterative Clustering}
\newacronym{gin}{GIN}{Graph Isomorphism Network} 
\newacronym{pna}{PNA}{Principal Neighboorhood Aggregation}
\newacronym{dgl}{DGL}{Deep Graph Library}

\newcommand{\ie}{\textit{i}.\textit{e}., }
\newcommand{\eg}{\textit{e}.\textit{g}., }

\newcommand{\gnnexplainer}{\textsc{GnnExplainer}}
\newcommand{\graphgradcam}{\textsc{GraphGrad-CAM}}
\newcommand{\graphgradcampp}{\textsc{GraphGrad-CAM++}}

\newcommand{\graphlrp}{\textsc{GraphLRP}}

\newcommand{\xmark}{\ding{55}}%
\newcommand{\cmark}{\ding{51}}%

\newcommand{\histocartography}{\textsc{HistoCartography}}
\newcommand{\histolab}{\textsc{Histolab}}
\newcommand{\staintools}{\textsc{StainTools}}
\newcommand{\qupath}{\textsc{QuPath}}
\newcommand{\syntax}{\textsc{Syntax}}
\newcommand{\openslide}{\textsc{OpenSlide}}
\newcommand{\histomistk}{\textsc{HistomicsTK}}

\newcommand{\image}{\mathtt{I}}
\newcommand{\tensor}{\mathtt{X}}
\newcommand{\mask}{\mathtt{M}}
\newcommand{\logits}{\mathtt{P}}
\newcommand{\importance}{\mathtt{S}}
\newcommand{\graph}{\mathtt{G}}

\begin{document}

\title{HistoCartography: A Toolkit for Graph Analytics in\\ Digital Pathology}

\author{\name Guillaume Jaume* \email gja@zurich.ibm.com \\
      \addr IBM Research, Zurich\\
      EPFL, Lausanne
      \AND
      \name Pushpak Pati* \email pus@zurich.ibm.com \\
      \addr IBM Research, Zurich\\
      ETH, Zurich
      \AND
      \name Valentin Anklin \email anklinv@student.ethz.ch \\
      \addr ETH, Zurich
      \AND
      \name Antonio Foncubierta \email fra@zurich.ibm.com \\
      \addr IBM Research, Zurich
      \AND
      \name Maria Gabrani \email mga@zurich.ibm.com \\
      \addr IBM Research, Zurich
}


\maketitle

\begin{abstract}
Advances in entity-graph based analysis of histopathology images have brought in a new paradigm to describe tissue composition, and learn the tissue structure-to-function relationship.
Entity-graphs offer flexible and scalable representations to characterize tissue organization, while allowing the incorporation of prior pathological knowledge to further support model interpretability and explainability.
However, entity-graph analysis requires prerequisites for image-to-graph translation and knowledge of state-of-the-art machine learning algorithms applied to graph-structured data, which can potentially hinder their adoption.
In this work, we aim to alleviate these issues by developing $\histocartography$, a standardized python API with necessary preprocessing, machine learning and explainability tools to facilitate graph-analytics in computational pathology. Further, we have benchmarked the computational time and performance on multiple datasets across different imaging types and histopathology tasks to highlight the applicability of the API for building computational pathology workflows.
\end{abstract}


\section{Introduction}

Recent advancements in tissue-slide digitization have paved way for enhancing storage, sharing capabilities, and computer-aided inspection by leveraging \gls{dl}.
Most \gls{dl} approaches analyze tissue images in three steps, namely patch generation, patch-level feature extraction, and feature aggregation to produce image-level embeddings for downstream pathology tasks.
However, they suffer from several limitations,
\begin{inparaenum}[(i)]
    \item the trade-off between operational resolution and adequate context per-patch,
    \item the aggregation is often sub-optimal,
    \item comprehensive modeling of tissue composition is missing, and
    \item the lack of model transparency raises barriers to deployment in real life.
\end{inparaenum}  

To circumvent these limitations, entity-graphs are proposed~\citep{demir04} where the nodes and edges of the graphs denote tissue entities and their interactions, respectively.
Entity-graphs, followed by \glspl{gnn}-based processing, have recently gained popularity in addressing various pathology tasks~\citep{zhou19,chen20,pati21,anklin21,jaume21}.
The entities can be biologically-defined, \eg nuclei, tissue regions, glands~\citep{zhou19,pati21,anklin21}, or can be patches~\citep{adnan20,aygunes20}.
The entity-graphs enable to simultaneously capture local entity environment and global tissue composition. They can seamlessly scale to arbitrary tissue dimensions by incorporating arbitrary number of entities and interactions, thus offering an alternate to \gls{mil}~\citep{campanella19,lu21}.
The entity-graphs also enable to selectively operate on diagnostically relevant entities, instead of analyzing the entire tissue~\citep{tellez20, shaban20}.
Furthermore, when the entities depict biological units, such as nuclei, glands etc., the analysis allows pathologists to directly comprehend and reason with the outcomes~\citep{jaume20, jaume21}. 
However, constructing an entity-graph based pathological workflow demands several prerequisites, such as entity detection, entity encoding, constructing the graph topology etc., alongside  standard preprocessing, such as stain normalization, tissue detection etc. Additionally, the workflow requires to utilize the recent advancements in \gls{dl} for processing graph-structured data. All these may inhibit the adoption of entity-graphs in computational pathology. In addition, the lack of a standardized framework with the aforementioned functionalities urge the researchers to reinvent the wheel, which is cumbersome, time-consuming, hampers reproducibility, and requires a wide range of technical acumen.

To overcome these constraints, we present $\histocartography$, a novel open-source python library that facilitates graph-analytics in computational pathology.
Specifically our contributions are:
\begin{inparaenum}[(i)]
    \item a standardized python library that unifies a set of histology image manipulation tools, entity-graph builders, \gls{gnn} models, and model explainability tools,
    \item a benchmark assessment of performance and scalability on classification and segmentation tasks in pathology,
    \item a comprehensive overview of graph representation and modeling in histology, and
    \item a review of extant libraries for histological image analysis.
\end{inparaenum}

\section{Related Work}

\subsection{Graphs in Computational Pathology}
Entity-graphs are proposed to realize the tissue composition-to-functionality relationship in terms of the phenotypical and structural characteristics of tissue. 
The entities can be nuclei~\citep{demir04,zhou19,wang20b,chen20,pati21}, tissue regions~\citep{pati21}, patches~\citep{anand20,adnan20,aygunes20,zhao20,li18,levy21}, etc. 
Typically nodes include handcrafted or \gls{dl} features to characterize the entities, and the topology can depict the spatial or semantic relationship among the entities, \eg \gls{knn}, region adjacency, or probabilistic models.
The graphs can be processed using classic \gls{ml}~\citep{sharma16,sharma17} or \glspl{gnn} to outperform state-of-the-art \gls{cnn}-based approaches for several pathology tasks across multiple organs~\citep{arteaga17,zhou19, zhao20,adnan20,pati21, studer21,anklin21}. 
Interestingly, when the graph-nodes depict biological entities, \eg nuclei, tissue regions, the entity-graphs combined with feature attribution techniques can provide pathologist-friendly interpretations~\citep{zhou19,jaume20,sureka20} and explanations~\citep{jaume21}, unlike pixelated blurry saliency maps. A detailed review of graphs in computational pathology is presented by~\cite{ahmedt21}.

\subsection{Extant Libraries in Computational Pathology}
Several open-source libraries facilitate the development of computational pathology workflows. Most of them include helper functions to perform standard preprocessing and visualization.
$\histolab$~\citep{histolab20} includes \gls{wsi}-level tissue detection and tile extraction modules. 
$\syntax$~\citep{syntax20} provides the same features with abstraction where modules can be stacked and run in a pre-defined pipeline.
$\staintools$~\citep{staintools19} provides tools for stain normalization and augmentation.
$\histomistk$~\citep{histomicstk21} enables to perform tissue detection, object detection and segmentation, image filtering, stain normalization and deconvolution, and handcrafted feature extraction. Further, $\histomistk$ allows nuclei segmentation and classification using classical \gls{ml} approaches. It also provides a \gls{ui} to run containerized modules and pipelines. Though $\histomistk$ includes valuable functionalities, it caters limited \gls{dl} tools. 
Similarly, $\openslide$~\citep{openslide20} provides a \gls{ui} to read and visualize histology images that supports most of the \gls{wsi} formats. 
Finally, $\qupath$~\citep{qupath21} offers a \gls{ui} that allows to read, visualize and annotate \glspl{wsi}. It also includes tools to perform stain normalization, nuclei and tissue detection, and implement basic \gls{ml} models. However, $\qupath$ is not a python \gls{api}, which makes it difficult to integrate into existing workflow and \gls{dl} frameworks, \eg PyTorch, Tensorflow.
Most importantly, none of the frameworks provide graph-related helpers. 
With the advent of graph-techniques as a new paradigm for analyzing histology images, a standardized library is desired for reinforcing the development.

\section{Histocartography: Graph Analytics Tool for Pathology}
In this section, we highlight the core functionalities of $\histocartography$, 
\begin{inparaenum}[(1)]
    \item \emph{Preprocessing} module: a set of histology image processing tools and entity-graph builders,
    \item \emph{\gls{ml}} module: helpers to learn from entity-graphs,
    \item \emph{Explainability} module: a set of \gls{gnn} model interpretability tools.
\end{inparaenum}
The specific functionalities in each module are summarized in Table~\ref{tab:overview}.

\begin{table*}[t]
\caption{Overview of $\histocartography$ functionalities, with the i/o, CPU and GPU compatibility, and availability in extant libraries for individual module.
$\image$, $\mask$, $\tensor$, $\graph$, $\logits$ and $\importance$ denote an image (np.array~\citep{harris2020}), a mask (np.array), features (torch.Tensor~\citep{paszke19}), a graph (DGLGraph~\citep{wang19}), predictions (torch.Tensor) and importance scores (torch.Tensor), respectively.}
\label{tab:overview}
\centering
\scriptsize 
\begin{tabular}{c|lccccc}
  \toprule
  Function & Module & Input & Output & Existing & CPU & GPU \\
  \midrule
  
  \parbox[t]{2mm}{\multirow{11}{*}[-1.4ex]{\rotatebox[origin=c]{90}{Preprocessing}}} & Vahadane Stain Norm & $\image$ & $\image$ & \cmark & \cmark & \xmark \\ [0.1cm]

  & Macenko Stain Norm & $\image$ & $\image$ & \cmark & \cmark & \xmark \\ [0.1cm]

  & Tissue Mask Detection & $\image$ & $\mask$ & \cmark & \cmark & \xmark \\ [0.1cm]

  & Nuclei Detection & $\image$ & $\mask$ &  \cmark & \cmark & \cmark \\ [0.1cm]

  & Nuclei Concepts & $\image$, $\mask$ & $\mask$ & \cmark & \cmark & \xmark \\ [0.1cm]

  & Tissue Component Detection & $\image$ & $\mask$ & \xmark & \cmark & \xmark \\ [0.1cm]

  & Deep Feature Extraction & $\image$, $\mask$ & $\tensor$ & \xmark & \cmark & \cmark \\ [0.1cm]

  & Feature Cube Extraction & $\image$ & $\tensor$ & \xmark & \cmark & \cmark \\ [0.1cm]

  & \gls{knn} Graph Building & $\tensor$, $\mask$  & $\graph$ & \xmark & \cmark & \xmark \\ [0.1cm]

  & RAG Graph Building & $\tensor$, $\mask$ & $\graph$ & \xmark & \cmark & \xmark \\ [0.1cm]

  \midrule
  \parbox[t]{2mm}{\multirow{3}{*}[-0.8ex]{\rotatebox[origin=c]{90}{ML}}} & Cell-Graph Model & $\graph$ & $\logits$ & \xmark & \cmark & \cmark \\ [0.1cm]

  & Tissue-Graph Model & $\graph$ & $\logits$ & \xmark & \cmark & \cmark \\ [0.1cm]  

  & HACT Model & $\graph$, $\graph$, $\tensor$  & $\logits$ & \xmark & \cmark & \cmark \\ [0.1cm]

  \midrule
  
  \parbox[t]{2mm}{\multirow{4}{*}[-0.8ex]{\rotatebox[origin=c]{90}{Explainers}}} & $\gnnexplainer$ & $\graph$ & $\importance$ & \xmark & \cmark & \cmark \\ [0.1cm]

  & $\graphgradcam$ & $\graph$ & $\importance$ & \xmark & \cmark & \cmark \\ [0.1cm]

  & $\graphgradcampp$ & $\graph$ & $\importance$ & \xmark & \cmark & \cmark \\ [0.1cm]

  & $\graphlrp$ & $\graph$ & $\importance$ & \xmark & \cmark & \cmark \\ [0.1cm]

  \bottomrule
\end{tabular}
\vspace{-1em}
\end{table*}

\subsection{Preprocessing Module}

\begin{figure}[!t]
\centering
\centerline{\includegraphics[width=0.98\textwidth]{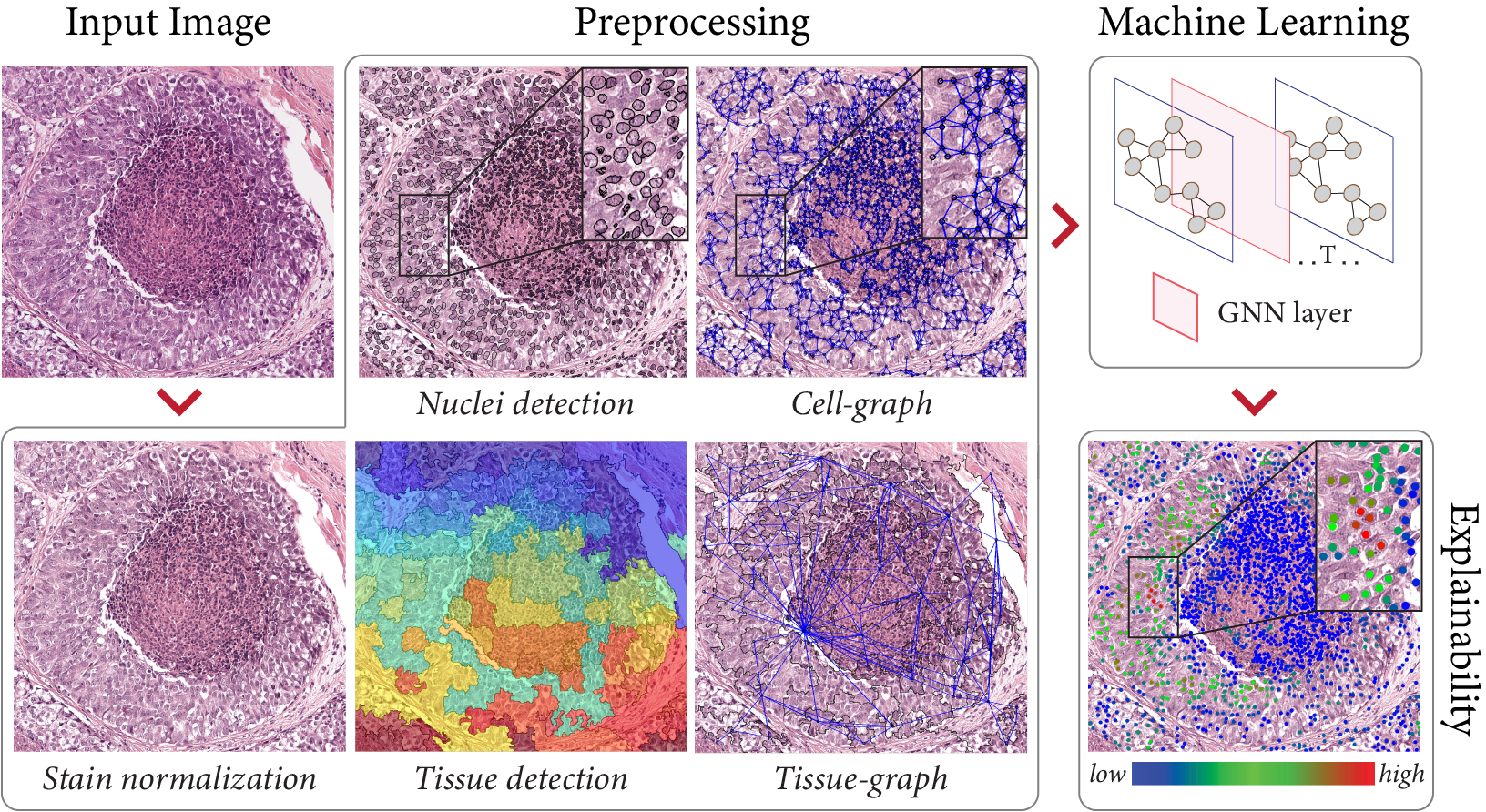}}
\vspace{-0.5em}
\caption{Overview of $\histocartography$ functionalities and modules.}
\label{fig:overview}
\vspace{-1em}
\end{figure}

\paragraph{Stain normalization:}
Variation in \gls{he} staining protocols for tissue specimens induces appearance variability that adversely impacts computational methods~\citep{tellez19}. 
To alleviate these variations, $\histocartography$ implements two popular normalization algorithms proposed by~\cite{macenko09} and~\cite{vahadane2016}, similar to $\staintools$ and $\histomistk$, which supports both reference-based and reference-free normalization, \ie with manual stain vectors. Figure~\ref{fig:overview} highlights a sample normalization output using our \gls{api}. 

\paragraph{Tissue Detection:}
A \gls{wsi} usually includes significant non-tissue region. Identifying the tissue regions can confine the analysis and reduce computational effort.
The tissue detector in $\histocartography$ iteratively applies Gaussian smoothing and Otsu thresholding until the mean of non-tissue pixels is below a threshold. 
This module is common across $\histolab$, $\syntax$, $\histomistk$ and $\qupath$. 

\paragraph{Nuclei detection:}
This module enables to segment and locate nuclei in \gls{he} images.
Though it is well-studied in computational pathology, only a few public implementations are  available. 
For instance, $\qupath$ allows to detect nuclei but requires model training and fine-tuning. While providing flexibility, the module includes only elementary \gls{ml} methods.
$\histocartography$ integrates two checkpoints for the state-of-the-art HoVerNet model~\citep{graham19} trained on PanNuke~\citep{gamper2020pannuke} and MoNuSac~\citep{monusac2020} datasets for nuclei segmentation and classification. 

\paragraph{Tissue Component Detection:}
$\histocartography$ includes an unsupervised superpixel based approach to segment tissue regions. First, the tissue is oversegmented into homogeneous superpixels using \gls{slic}~\citep{achanta12} algorithm. Then, neighboring superpixels are hierarchically merged using color similarity to denote meaningful tissue regions, \eg epithelium and stroma regions. 
Superpixels depicting tissue regions are used by~\cite{bejnordi15,pati20, pati21}.

\paragraph{Feature Extraction:}
$\histocartography$ includes two types of feature extractors, \ie handcrafted- and \gls{cnn}-based, to encode the entity characteristics. 
The handcrafted feature extractor computes entity-level morphological and topological properties. Morphological features capture the shape and size, \eg entity area, eccentricity, perimeter etc., and the texture captures chromaticity using the gray-level co-occurrence matrix. 
Topological features capture the local entity distribution using \gls{knn} entity density estimation.
A comprehensive list is provided in the Appendix. Handcrafted features can be used for training \gls{dl} algorithms~\citep{demir04,zhou19,pati20,studer21}, or concept-based post-hoc explainability~\citep{jaume21}. 

The deep feature extractor allows to extract \gls{cnn} features by using any pre-trained deep architecture, \eg ResNet, MobileNet, embedded in torchvision~\citep{marcel10}. The module intakes patches centered around the entity and extracts features from the penultimate layer of the architectures. 
If the entity is larger than the specified patch size, then multiple patches within the entity, w/ or w/o overlapping, are processed, and the final feature is computed as the mean of the patch-level deep features, as used in~\cite{chen20,pati20,pati21}. 
Deep features can alternatively be extracted from the \gls{wsi} to build a feature-cube as suggested by~\cite{shaban20,tellez20}.
\vspace{-0.25em}

\paragraph{Graph builders:}
$\histocartography$ presents two graph builders, \ie the thresholded \gls{knn} and the \gls{rag}.
The \gls{knn} graph builder defines the graph topology by connecting each entity to its k-closest neighbors. Connections between distant entities beyond a threshold can be pruned to have spatial sparsity in the graph. We recommend this builder to connect single entities, \eg nuclei, glands.
The \gls{rag} builder connects entities using spatial adjacency, \ie entities sharing a common boundary. It builds a sound topology when dealing with dense segmentation maps, \eg tissue regions. Figure~\ref{fig:overview} presents samples of cell- and tissue-graphs.
Further, the module fuses the node features and the topological distribution to render a \gls{dgl} graph for an image.

\subsection{Graph Machine Learning Module}
$\histocartography$ includes a set of \gls{dl} models, based on a \gls{gnn} backbone to learn from graph-structured tissue representations. 
It includes two state-of-the-art \gls{gnn} layers, \ie \gls{gin}~\citep{xu19} and \gls{pna}~\citep{corso20}. \gls{pna} proves to outperform \gls{gin} provided more computational resources~\citep{dwivedi20}.
$\histocartography$ defines cell- and tissue-graph models, which are \gls{gnn}-based abstractions to learn from biological entity-graphs. They offer efficient~\citep{pati21}, scalable~\citep{anklin21} and explainable~\citep{jaume20, jaume21} approaches to analyze histology images. Further, the library includes models to jointly represent and learn from cell- and tissue-graphs~\citep{pati21}. 
The models in $\histocartography$ are organized such that they can be adapted to various \gls{gnn} backbones, tasks (\eg regression, clustering, classification, segmentation), organs, and entity-types. These models provide the blueprints to accelerate the development of graph-based models in computational pathology. All the graph modules are implemented using \gls{dgl}~\citep{wang19}, a state-of-the-art library for \glspl{gnn} built around PyTorch.

\subsection{Explainability Module}
$\histocartography$ includes four post-hoc feature attribution graph explainers, that can generate node-level saliency maps to highlight the node-wise contribution towards an output task.
Namely, the library includes two gradient-based explainers~($\graphgradcam$~\citep{selvaraju17, pope19} and $\graphgradcampp$~\citep{chattopadhay18, jaume21}), a node pruning-based explainer ($\gnnexplainer$~\citep{ying19}), and a layer-wise relevance propagation explainer~($\graphlrp$~\citep{schwarzenberg19}).
The saliency map can be visualized by overlaying the node importances on the input image (see Figure~\ref{fig:explanation}). Alternatively, entities with high importances can be extracted and studied independently to assess their relevance~\citep{jaume21}.

\subsection{Pipeline Runner}
To facilitate an easy-to-use and human-readable development, $\histocartography$ includes a pipeline runner. It allows to define a series of pipeline steps along with loading and saving utilities to reduce boilerplate code.

\section{Benchmarking $\histocartography$}
We benchmark $\histocartography$ in terms of run-time and performance for various histopathology tasks, \ie stain normalization, tissue detection, tumor classification and segmentation etc., on images of varying dimensions. The CPU and GPU compatible modules are assessed on a single-core POWER8 processor and a NVIDIA P100 GPU, respectively.

\begin{table}[t]
\caption{Run time analysis of $\histocartography$ core functionalities (in seconds).}
\label{tab:time}
\centering
\scriptsize 
\begin{tabular}{c|l|ccc|ccc}
  \toprule
  
  Functionalities & Modules & \multicolumn{3}{c|}{Tumor RoI} & \multicolumn{3}{c}{\gls{wsi}} \\
  \midrule
  
  & Size & $1000^2$ & $2500^2$ & $5000^2$ & $5000^2$ & $7500^2$ & $11000^2$ \\
  \midrule

  \parbox[t]{2mm}{\multirow{6}{*}[-1.4ex]{\rotatebox[origin=c]{90}{Preprocessing}}} & Vahadane Stain Norm & $1.77$ & $6.46$ & $29.03$ & $30.67$ & $68.27$ & $186.10$ \\ [0.1cm]

  & Macenko Stain Norm & $0.80$ & $2.86$ & $11.19$ & $15.98$ & $32.37$ & $81.72$ \\ [0.1cm]

  & Tissue Mask Detection & - & - & - & $1.04$ & $2.11$ & $8.09$ \\ [0.1cm]

  & Feature Cube Extraction & $0.24$ & $1.61$ & $5.92$ & $5.83$ & $11.77$ & $30.21$ \\ [0.1cm]

  & Cell-Graph Generation & $2.51$ & $13.33$ & $50.26$ & - & - & - \\ [0.1cm]

  & Tissue-Graph Generation & $4.14$ & $21.30$ & $83.66$ & $39.18$ & $93.26$ & $276.71$ \\ [0.1cm]

  \midrule
  
  \parbox[t]{2mm}{\multirow{3}{*}[-0.8ex]{\rotatebox[origin=c]{90}{ML}}} & Cell-Graph Model & $0.028$ & $0.033$ & $0.040$ & - & - & - \\ [0.1cm]

  & Tissue-Graph Model & $0.011$ & $0.015$ & $0.026$ & $0.039$ & $0.056$ & $0.069$ \\ [0.1cm]  

  & Hierarchical Model & $0.034$ & $0.041$ & $0.057$ & - & - & - \\ [0.1cm]
  \midrule
  
  \parbox[t]{2mm}{\multirow{4}{*}[-0.8ex]{\rotatebox[origin=c]{90}{Explainers}}} & $\gnnexplainer$ & $12.00$ & $13.09$ & $35.33$ & - & - & - \\ [0.1cm]

  & $\graphgradcam$ & $0.011$ & $0.022$ & $0.035$ & $0.025$ & $0.030$ & $0.033$ \\ [0.1cm]

  & $\graphgradcampp$ & $0.011$ & $0.023$ & $0.035$ & $0.026$ & $0.030$ & $0.033$ \\ [0.1cm]

  & $\graphlrp$ & $0.020$ & $0.024$ & $0.90$ & $0.079$ & $0.085$ & $0.089$ \\ [0.1cm]
  \bottomrule
\end{tabular}
\vspace{-1em}
\end{table}

\subsection{Computational Time}
Analyzing the computational time for processing a histology image is imperative. 
We thoroughly analyze the run-time of $\histocartography$ modules on a set of \glspl{roi} and \glspl{wsi}. The analyses are presented in Table~\ref{tab:time}, and a comprehensive extension is provided in Supplementary: Table~\ref{tab:extended_time}. 
The preprocessing modules are observed to be the most time-consuming. 
For instance, Vahadane stain normalization can take up to 3 minutes to process a $11'000 \times 11'000$ image, whereas Macenko method is 2$\times$ faster for competitive result. The implementations are computationally similar to $\histolab$ and $\staintools$, and scale linearly w.r.to image size.
The cell- and tissue-graph construction take 2.5 and 4.1 seconds respectively for a $1000\times1000$ image with the following parameters. Nuclei detection is performed on patches of size $256\times256$ with an overlap of $164$ pixels. 
Nuclei features are extracted from $72\times72$ patches centered around the nuclei, that are resized to $224\times224$ and processed by ResNet34 pretrained on ImageNet~\citep{deng09}. Finally, thresholded k-NN topology is built with $k=5$ and a threshold distance of 50 pixels.
For the tissue-graph, \gls{slic} is used to extract 400 superpixels per image, that are subsequently merged to provide the tissue components. Tissue features are also extracted using ResNet34 with $144\times144$ size patches that are resized to $224 \times 224$. 
The graph buildings can be further optimized as per the task by downsampling the input image, reducing the patch overlap, or by using a lighter feature extractor. 
For extracting the feature cube representation, we process patches of size $144\times144$ resized to $224\times224$ w/o overlap by pretrained ResNet34. 

\glspl{troi} are processed using a cell- and tissue-graph model, and the hierarchical cell-to-tissue graph model~\citep{pati21}. They consist of three PNA layers with 64 hidden units followed by an additional 2-layer MLP with 128 hidden units for classification. 
WSIs are processed using \textsc{SegGini}~\cite{anklin21}, a weakly supervised approach basdn on tissue-graphs, which contains six GIN layers with 64 hidden units followed by a 2-layer MLP with 128 hidden units. 
The models process in near real-time irrespective of the increment in the graph size. 
The graph explainers are based on GNNs with 3 GIN layers, each having a 2-layer MLP with 32 hidden units, and a 2-layer MLP head. $\gnnexplainer$ is the slowest among all as it requires to optimize a mask to explain each image.

\begin{table}[t]
\caption{Benchmarking $\histocartography$ for classification and segmentation (in \%).}
\label{tab:performance}
\centering
\scriptsize 
\begin{tabular}{c|lcccccc}
  \toprule
  
  Task & Dataset & Model & Image Type & Avg. \#pixels & \#classes & Avg. Dice & Weighted F1 \\
  \midrule
  
  \parbox[t]{2mm}{\multirow{6}{*}[-1.4ex]{\rotatebox[origin=c]{90}{Classification}}} & BRACS & \textsc{CG-GNN} & \gls{troi} & $3.9 \times 10^6$ (40$\times$) & 7 & - & $55.9\pm1.0$  \\ [0.1cm]
  
  & BRACS & \textsc{TG-GNN} & \gls{troi} & $3.9 \times 10^6$ (40$\times$) & 7 & - & $56.6\pm1.3$  \\ [0.1cm]
  
  & BRACS & \textsc{HACT}-Net & \gls{troi} & $3.9 \times 10^6$ (40$\times$) & 7 & - & $61.5\pm0.9$  \\ [0.1cm]

  & BACH & \textsc{HACT}-Net & \gls{troi} & $3.1 \times 10^6$ (20$\times$) & 4 & - & $90.7\pm0.5$  \\ [0.1cm]
  
  & SICAPv2 & \textsc{SegGini} & \gls{wsi} & $121 \times 10^6$ (10$\times$) & 6 & - & $62.0\pm3.6$  \\ [0.1cm]

  & UZH & \textsc{SegGini} & TMA & $9.6 \times 10^6$ (40$\times$) & 6 & - & $56.8\pm1.7$  \\ [0.1cm]
  \midrule

  \parbox[t]{2mm}{\multirow{2}{*}[-0.8ex]{\rotatebox[origin=c]{90}{Seg.}}} & SICAPv2 & \textsc{SegGini} & \gls{wsi} & $121 \times 10^6$ (10$\times$) & 4 & $44.3\pm2.0$ & - \\ [0.1cm]

  & UZH & \textsc{SegGini} & TMA & $9.6 \times 10^6$ (40$\times$) & 4 & $66.0\pm3.1$ & - \\ [0.1cm] 
  \bottomrule
\end{tabular}
\end{table}

\begin{figure}[t]
\centering
\centerline{\includegraphics[width=0.98\textwidth]{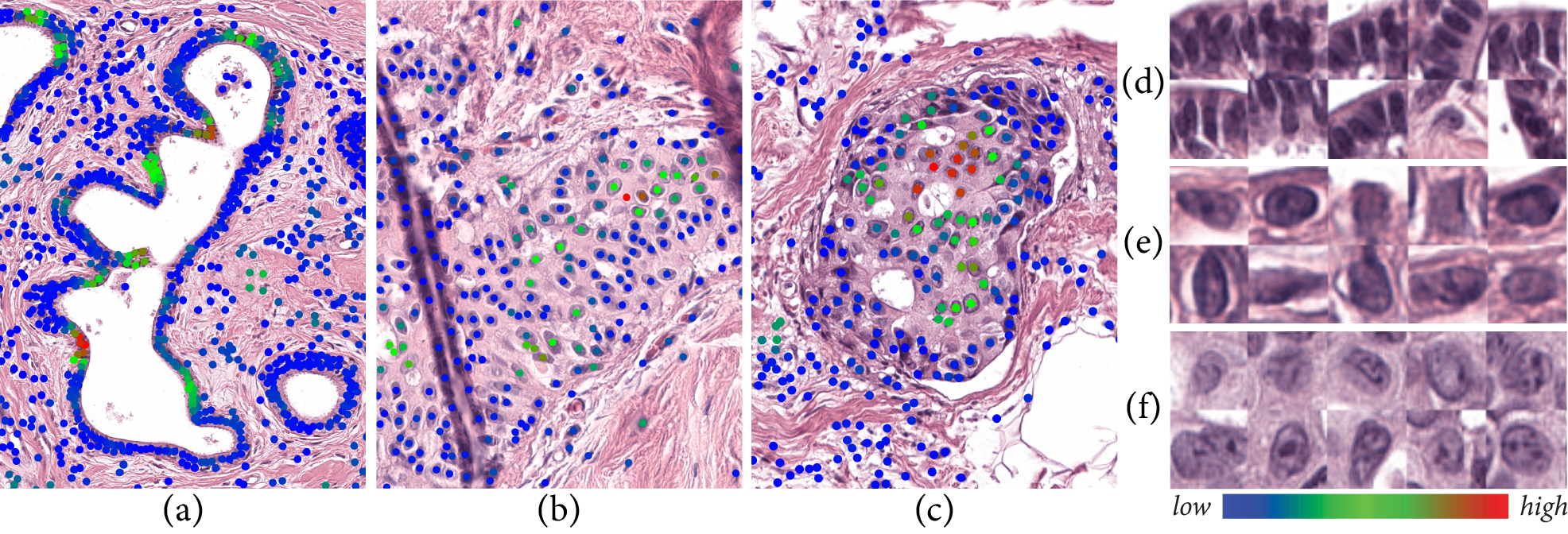}}
\vspace{-1em}
\caption{Qualitative explanations of sample breast \glspl{roi}: (a) Benign, (b) ADH, (c) DCIS. (d, e, f) highlight the ten most important nuclei for the respective samples.}
\label{fig:explanation}
\vspace{-2em}
\end{figure}

\subsection{Performance Benchmark}
Table~\ref{tab:performance} benchmarks the performance of $\histocartography$ for classification and segmentation tasks. 
Classification is performed on BRACS~\citep{pati21} and BACH~\citep{aresta19} datasets to characterize breast tumors using cell-graph model, tissue-graph model, and HACT-Net~\citep{pati21}, and the performance is measures by weighted-F1 score.
Segmentation is performed using \textsc{SegGini}~\citep{anklin21} to delineate Gleason patterns in prostate cancer images from UZH~\citep{zhong2017} and SICAPv2~\citep{silva2020}, and the performance is measured by average Dice score.
We evaluate on various image types, \ie tumor \glspl{roi}, tissue microarrays, and whole-slides, to highlight the scalability of entity-graphs in $\histocartography$ to arbitrary image dimensions.

\subsection{Qualitative Explanations}
Figure~\ref{fig:explanation} presents the outcome of \textsc{GraphGradCAM} module in $\histocartography$ to interpret a cell-graph model. This module renders per-image explanations in terms of node-level saliency maps by applying post-hoc feature attribution methods on trained cell-graph model. Further, the cell-graph model can be interpreted by characterizing the highlighted important nuclei per-image, as shown in Figure~\ref{fig:explanation}.

\vspace{-0.75em}
\section{Conclusion}
We introduced $\histocartography$, the first open source library, to the best of our knowledge, to facilitate graph analytics, \ie graph representation, learning, and explainability, in computational pathology. 
It can potentially enable researchers to develop entity-graph based pathology workflows by leveraging the inbuilt helpers. 
As the library is built on python, the deep learning researchers can seamlessly customize and integrate the functionalities into their task-specific workflows.
$\histocartography$ is constantly growing with new functionalities and improved implementations, aiming to promote the adoption of graph-based analysis in computational pathology.

\bibliography{main}

\begin{thebibliography}{55}
\providecommand{\natexlab}[1]{#1}
\providecommand{\url}[1]{\texttt{#1}}
\expandafter\ifx\csname urlstyle\endcsname\relax
  \providecommand{\doi}[1]{doi: #1}\else
  \providecommand{\doi}{doi: \begingroup \urlstyle{rm}\Url}\fi

\bibitem[Achanta et~al.(2012)]{achanta12}
R.~Achanta et~al.
\newblock Slic superpixels compared to state-of-the-art superpixel methods.
\newblock In \emph{IEEE Transactions on Pattern Analysis and Machine
  Intelligence}, volume~34, pages 2274--2282, 2012.

\bibitem[Adnan et~al.(2020)]{adnan20}
M.~Adnan et~al.
\newblock Representation learning of histopathology images using graph neural
  networks.
\newblock In \emph{IEEE Conference on Computer Vision and Pattern Recognition
  (CVPR) Workshops}, pages 4254--4261, 2020.

\bibitem[Ahmedt-Aristizabal et~al.(2021)]{ahmedt21}
D.~Ahmedt-Aristizabal et~al.
\newblock A survey on graph-based deep learningfor computational
  histopathology.
\newblock In \emph{arXiv:2107.00272}, 2021.

\bibitem[Anand et~al.(2020)]{anand20}
D.~Anand et~al.
\newblock Histographs: graphs in histopathology.
\newblock In \emph{SPIE Medical Imaging 2020: Digital Pathology}, volume 11320,
  page 113200O, 2020.

\bibitem[Anklin et~al.(2021)]{anklin21}
V.~Anklin et~al.
\newblock Learning whole-slide segmentation from inexact and incomplete labels
  using tissue graphs.
\newblock 2021.

\bibitem[Arbitrio et~al.()]{histolab20}
E.~Arbitrio et~al.
\newblock histolab.
\newblock URL \url{https://github.com/histolab/histolab}.

\bibitem[Aresta et~al.(2019)]{aresta19}
G.~Aresta et~al.
\newblock Bach: Grand challenge on breast cancer histology images.
\newblock In \emph{Medical Image Analysis}, volume~56, pages 122--139, 2019.

\bibitem[Aygunes et~al.(2020)]{aygunes20}
B.~Aygunes et~al.
\newblock Graph convolutional networks for region of interest classification in
  breast histopathology.
\newblock In \emph{SPIE Medical Imaging 2020: Digital Pathology}, volume 11320,
  page 113200K, 2020.

\bibitem[Bankhead et~al.()]{qupath21}
P.~Bankhead et~al.
\newblock Qupath.
\newblock URL \url{https://qupath.github.io/}.

\bibitem[Beezley et~al.()]{histomicstk21}
J.~Beezley et~al.
\newblock Histomicstk.
\newblock URL \url{https://github.com/DigitalSlideArchive/HistomicsTK}.

\bibitem[Bejnordi et~al.(2015)]{bejnordi15}
B.E. Bejnordi et~al.
\newblock A multi-scale superpixel classification approach to the detection of
  regions of interest in whole slide histopathology images.
\newblock In \emph{SPIE Medical Imaging}, 2015.

\bibitem[Byfield et~al.({\natexlab{a}})]{staintools19}
P.~Byfield et~al.
\newblock Staintools, {\natexlab{a}}.
\newblock URL \url{https://github.com/Peter554/StainTools}.

\bibitem[Byfield et~al.({\natexlab{b}})]{syntax20}
P.~Byfield et~al.
\newblock Syntax, {\natexlab{b}}.
\newblock URL \url{https://github.com/jgamper/compay-syntax/tags}.

\bibitem[Campanella et~al.(2019)]{campanella19}
G.~Campanella et~al.
\newblock Clinical-grade computational pathology using weakly supervised deep
  learning on whole slide images.
\newblock In \emph{Nature Medicine}, volume~25, page 1301–1309., 2019.

\bibitem[Chattopadhay et~al.(2018)]{chattopadhay18}
A.~Chattopadhay et~al.
\newblock {Grad-CAM++: Generalized gradient-based visual explanations for deep
  convolutional networks}.
\newblock In \emph{IEEE Winter Conference on Applications of Computer Vision},
  volume 2018-Janua, pages 839--847, 2018.

\bibitem[Chen et~al.(2020)]{chen20}
R.J. Chen et~al.
\newblock Pathomic fusion: An integrated framework for fusing histopathology
  and genomic features for cancer diagnosis and prognosis.
\newblock In \emph{IEEE Transactions on Medical Imaging}, 2020.

\bibitem[Corso et~al.(2020)]{corso20}
G.~Corso et~al.
\newblock Principal neighbourhood aggregation for graph nets.
\newblock In \emph{Neural Information Processing Systems (NeurIPS)}, 2020.

\bibitem[Demir et~al.(2004)]{demir04}
C.G. Demir et~al.
\newblock The cell graphs of cancer.
\newblock In \emph{Bioinformatics}, volume~20, pages 145--151, 2004.

\bibitem[Deng et~al.(2009)]{deng09}
J.~Deng et~al.
\newblock Imagenet: A large-scale hierarchical image database.
\newblock In \emph{IEEE Conference on Computer Vision and Pattern Recognition
  (CVPR)}, pages 248--255, 2009.

\bibitem[Dwivedi et~al.(2020)]{dwivedi20}
V.P. Dwivedi et~al.
\newblock Benchmarking graph neural networks.
\newblock In \emph{arXiv:2003.00982}, 2020.

\bibitem[Gamper et~al.(2020)]{gamper2020pannuke}
J.~Gamper et~al.
\newblock Pannuke dataset extension, insights and baselines.
\newblock \emph{arXiv preprint arXiv:2003.10778}, 2020.

\bibitem[Garciá-Arteaga et~al.(2017)]{arteaga17}
J.D. Garciá-Arteaga et~al.
\newblock A lymphocyte spatial distribution graph-based method for automated
  classification of recurrence risk on lung cancer images.
\newblock In \emph{International Symposium on Medical Information Processing
  and Analysis}, volume 10956, page 109560H, 2017.

\bibitem[Gilbert et~al.()]{openslide20}
B.~Gilbert et~al.
\newblock Openslide.
\newblock URL \url{https://github.com/openslide/openslide-python/}.

\bibitem[Graham et~al.(2019)]{graham19}
S.~Graham et~al.
\newblock Hover-net: Simultaneous segmentation and classification of nuclei in
  multi-tissue histology images.
\newblock In \emph{Medical Image Analysis}, volume~58, 2019.

\bibitem[Harris et~al.(2020)]{harris2020}
R.~P. Harris et~al.
\newblock Array programming with {NumPy}.
\newblock \emph{Nature}, 585\penalty0 (7825):\penalty0 357--362, September
  2020.
\newblock \doi{10.1038/s41586-020-2649-2}.
\newblock URL \url{https://doi.org/10.1038/s41586-020-2649-2}.

\bibitem[Jaume et~al.(2020)]{jaume20}
G.~Jaume et~al.
\newblock Towards explainable graph representations in digital pathology.
\newblock In \emph{International Conference on Machine Learning (ICML),
  Workshop on Computational Biology}, 2020.

\bibitem[Jaume et~al.(2021)]{jaume21}
G.~Jaume et~al.
\newblock Quantifying explainers of graph neural networks in computational
  pathology.
\newblock In \emph{IEEE Conference on Computer Vision and Pattern Recognition
  (CVPR)}, 2021.

\bibitem[Levy et~al.(2021)]{levy21}
J.~Levy et~al.
\newblock Topological feature extraction and visualization of whole slide
  images using graph neural networks.
\newblock \emph{Pacific Symposium on Biocomputing}, 26:\penalty0 285--296,
  2021.
\newblock \doi{10.1038/s41586-020-2649-2}.

\bibitem[Li et~al.(2018)]{li18}
R.~Li et~al.
\newblock Graph cnn for survival analysis on whole slide pathological images.
\newblock In Alejandro~F. Frangi, Julia~A. Schnabel, Christos Davatzikos,
  Carlos Alberola-L{\'o}pez, and Gabor Fichtinger, editors, \emph{Medical Image
  Computing and Computer Assisted Intervention -- MICCAI 2018}, pages 174--182,
  Cham, 2018. Springer International Publishing.

\bibitem[Macenko et~al.(2009)]{macenko09}
M.~Macenko et~al.
\newblock A method for normalizing histology slides for quantitative analysis.
\newblock In \emph{IEEE International Symposium on Biomedical Imaging (ISBI)},
  pages 1107--1110, 2009.

\bibitem[Marcel et~al.(2010)]{marcel10}
S.~Marcel et~al.
\newblock Torchvision the machine-vision package of torch.
\newblock In \emph{Proceedings of the 18th ACM International Conference on
  Multimedia}, MM '10, page 1485–1488, New York, NY, USA, 2010. Association
  for Computing Machinery.
\newblock ISBN 9781605589336.
\newblock \doi{10.1145/1873951.1874254}.
\newblock URL \url{https://doi.org/10.1145/1873951.1874254}.

\bibitem[Ming et~al.(2021)]{lu21}
Y.~Ming et~al.
\newblock Data efficient and weakly supervised computational pathology on whole
  slide images.
\newblock In \emph{Nature Biomedical Engineering}, 2021.

\bibitem[Paszke et~al.(2019)]{paszke19}
A.~Paszke et~al.
\newblock Pytorch: An imperative style, high-performance deep learning library.
\newblock In \emph{Neural Information Processing Systems (NeurIPS)}, pages
  8024--8035, 2019.

\bibitem[Pati et~al.(2020)]{pati20}
P.~Pati et~al.
\newblock Hact-net: A hierarchical cell-to-tissue graph neural network for
  histopathological image classification.
\newblock In \emph{Medical Image Computing and Computer Assisted Intervention
  (MICCAI) Workshop on GRaphs in biomedicAl Image anaLysis}, 2020.

\bibitem[Pati et~al.(2021)]{pati21}
P.~Pati et~al.
\newblock Hierarchical graph representations in digital pathology.
\newblock In \emph{Medical Image Analysis}, 2021.

\bibitem[Pope et~al.(2019)]{pope19}
P.E. Pope et~al.
\newblock Explainability methods for graph convolutional neural networks.
\newblock In \emph{IEEE Conference on Computer Vision and Pattern Recognition
  (CVPR)}, pages 10764--10773, 2019.

\bibitem[Ruchika et~al.(2020)]{monusac2020}
G.~Ruchika et~al.
\newblock Multi-organ nuclei segmentation and classification challenge 2020.
\newblock 02 2020.
\newblock \doi{10.13140/RG.2.2.12290.02244/1}.
\newblock URL \url{http://rgdoi.net/10.13140/RG.2.2.12290.02244/1}.

\bibitem[Schwarzenberg et~al.(2019)]{schwarzenberg19}
R.~Schwarzenberg et~al.
\newblock {Layerwise relevance visualization in convolutional text graph
  classifiers}.
\newblock \emph{EMNLP Workshop}, pages 58--62, 2019.

\bibitem[Selvaraju et~al.(2017)]{selvaraju17}
R.R. Selvaraju et~al.
\newblock {Grad-CAM : Visual Explanations from Deep Networks}.
\newblock In \emph{International Conference on Computer Vision}, pages
  618--626, 2017.

\bibitem[Shaban et~al.(2020)]{shaban20}
M.~Shaban et~al.
\newblock Context-aware convolutional neural network for grading of colorectal
  cancer histology images.
\newblock In \emph{IEEE Transactions on Medical Imaging}, volume~39, pages 2395
  -- 2405, 2020.

\bibitem[Sharma et~al.(2016)]{sharma16}
H.~Sharma et~al.
\newblock Cell nuclei attributed relational graphs for efficient representation
  and classification of gastric cancer in digital histopathology.
\newblock In \emph{SPIE Medical Imaging: Digital Pathology}, volume 9791, 2016.

\bibitem[Sharma et~al.(2017)]{sharma17}
H.~Sharma et~al.
\newblock A comparative study of cell nuclei attributed relational graphs for
  knowledge description and categorization in histopathological gastric cancer
  whole slide images.
\newblock In \emph{IEEE Symposium on Computer-Based Medical Systems}, pages
  61--66, 2017.

\bibitem[Silva-Rodríguez et~al.(2020)]{silva2020}
J.~Silva-Rodríguez et~al.
\newblock Going deeper through the gleason scoring scale: An automatic
  end-to-end system for histology prostate grading and cribriform pattern
  detection.
\newblock In \emph{Computer Methods and Programs in Biomedicine}, volume 195,
  2020.

\bibitem[Studer et~al.(2021)]{studer21}
L.~Studer et~al.
\newblock Classification of intestinal gland cell-graphs using graph neural
  networks.
\newblock In \emph{International Conference on Pattern Recognition (ICPR)},
  2021.

\bibitem[Sureka et~al.(2020)]{sureka20}
M.~Sureka et~al.
\newblock Visualization for histopathology images using graph convolutional
  neural networks.
\newblock In \emph{arXiv:2006.09464}, 2020.

\bibitem[Tellez et~al.(2019{\natexlab{a}})]{tellez19}
D.~Tellez et~al.
\newblock Quantifying the effects of data augmentation and stain color
  normalization in convolutional neural networks for computational pathology.
\newblock In \emph{Medical Image Analysis}, volume~58, 2019{\natexlab{a}}.

\bibitem[Tellez et~al.(2019{\natexlab{b}})]{tellez20}
D.~Tellez et~al.
\newblock Neural image compression for gigapixel histopathology image analysis.
\newblock volume~58, 2019{\natexlab{b}}.

\bibitem[Vahadane et~al.(2016)]{vahadane2016}
A.~Vahadane et~al.
\newblock Structure-preserving color normalization and sparse stain separation
  for histological images.
\newblock In \emph{IEEE Transactions on Medical Imaging}, volume~35, pages
  1962--1971, 2016.

\bibitem[Wang et~al.(2020)]{wang20b}
J.~Wang et~al.
\newblock Weakly supervised prostate tma classification via graph convolutional
  networks.
\newblock In \emph{2020 IEEE 17th International Symposium on Biomedical Imaging
  (ISBI)}, pages 239--243, 2020.
\newblock \doi{10.1109/ISBI45749.2020.9098534}.

\bibitem[Wang et~al.(2019)]{wang19}
M.~Wang et~al.
\newblock Deep graph library: Towards efficient and scalable deep learning on
  graphs.
\newblock In \emph{CoRR}, volume abs/1909.01315, 2019.

\bibitem[Xu et~al.(2019)]{xu19}
K.~Xu et~al.
\newblock How powerful are graph neural networks?
\newblock In \emph{International Conference on Learning Representations
  (ICLR)}, 2019.

\bibitem[Ying et~al.(2019)]{ying19}
R.~Ying et~al.
\newblock {GNNExplainer: Generating Explanations for Graph Neural Networks}.
\newblock In \emph{Advances in Neural Information Processing Systems}, 2019.

\bibitem[Zhao et~al.(2020)]{zhao20}
Y.~Zhao et~al.
\newblock Predicting lymph node metastasis using histopathological images based
  on multiple instance learning with deep graph convolution.
\newblock In \emph{IEEE Conference on Computer Vision and Pattern Recognition
  (CVPR)}, pages 4837--4846, 2020.

\bibitem[Zhong et~al.(2017)]{zhong2017}
Q.~Zhong et~al.
\newblock A curated collection of tissue microarray images and clinical outcome
  data of prostate cancer patients.
\newblock In \emph{Scientific Data}, volume~4, 2017.

\bibitem[Zhou et~al.(2019)]{zhou19}
Y.~Zhou et~al.
\newblock {CGC}-net: Cell graph convolutional network for grading of colorectal
  cancer histology images.
\newblock In \emph{Proceedings of the IEEE International Conference on Computer
  Vision (ICCV) Workshops}, 2019.

\end{thebibliography}

\clearpage
\appendix
\renewcommand\thefigure{\thesection\arabic{figure}}   
\renewcommand\thetable{\thesection\arabic{table}}   
\section*{Supplementary}

\subsection*{\textsc{HistoCartography} Syntax}
\setcounter{figure}{0}
\setcounter{table}{0}

\begin{figure}[h]
\centering
\centerline{\includegraphics[width=0.98\textwidth]{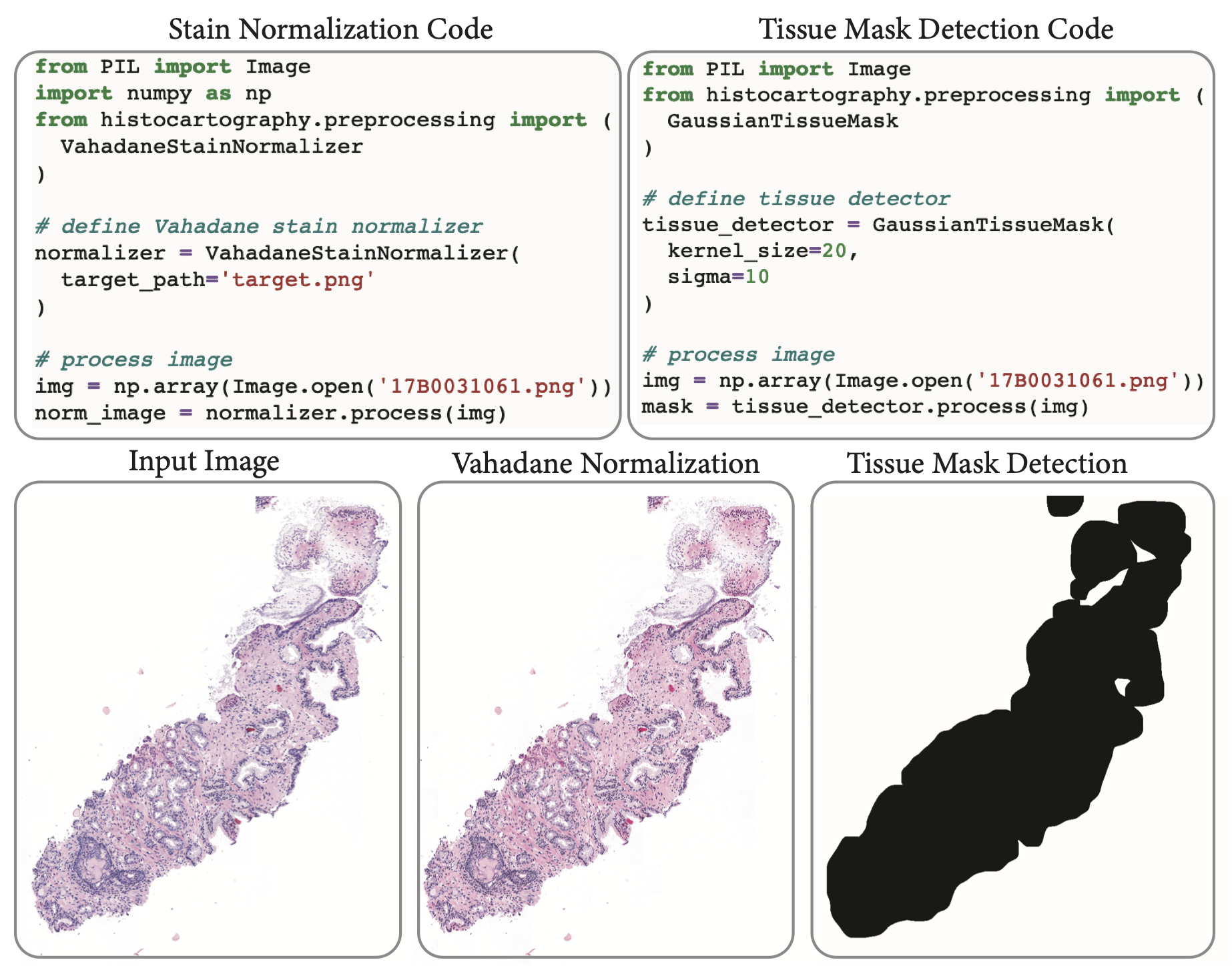}}
\caption{Implementation of Vahadane stain normalization (left) and tissue mask detection (right) with the \textit{Preprocessing} functionalities in the $\histocartography$ \gls{api}.}
\label{fig:code1}
\end{figure}

\begin{figure}[h]
\centering
\centerline{\includegraphics[width=0.98\textwidth]{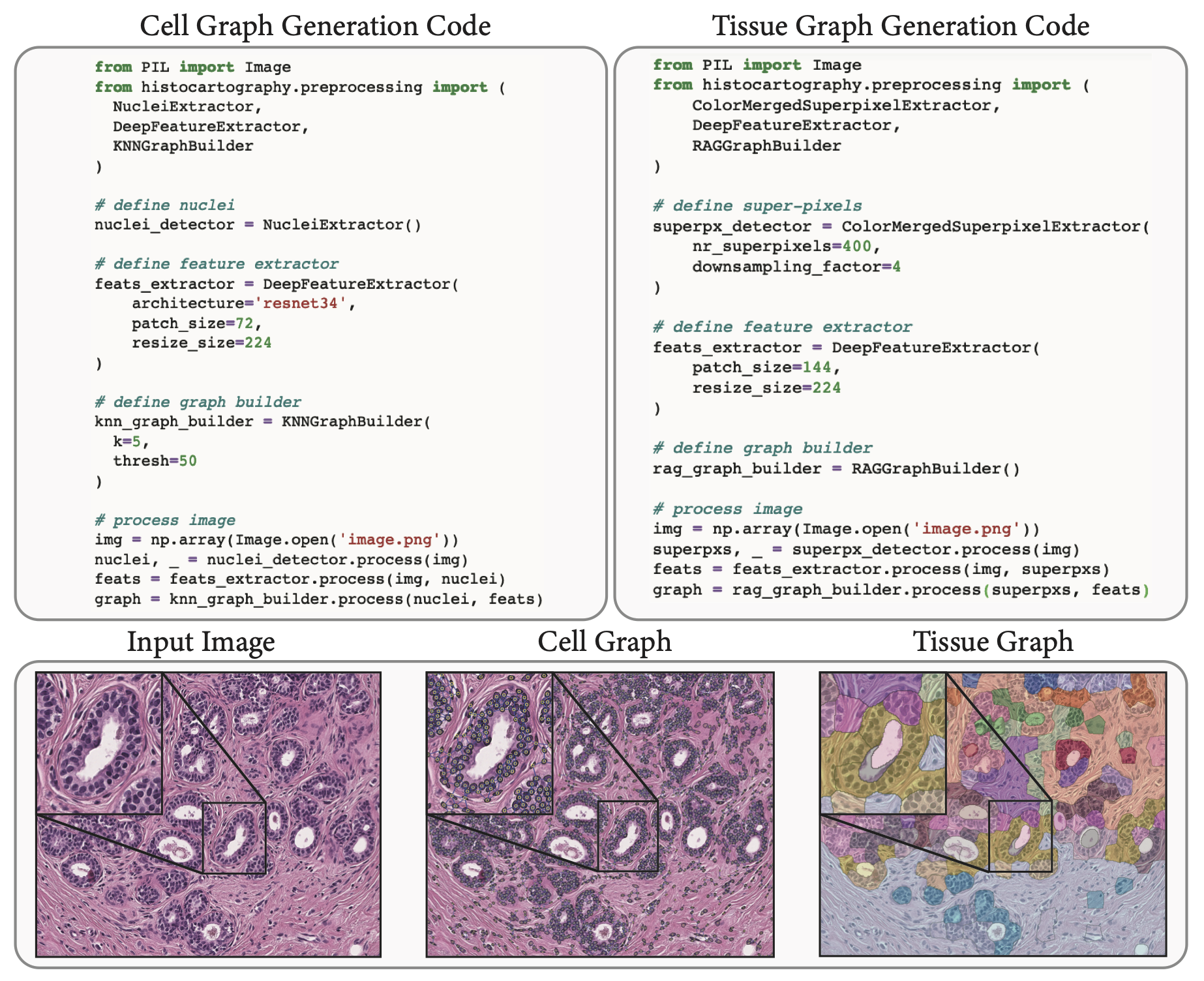}}
\caption{Implementation of cell-graph (left) and tissue-graph (right) generation using the graph builders in $\histocartography$.}
\label{fig:code2}
\vspace{-2em}
\end{figure}

\begin{figure}[t]
\centering
\centerline{\includegraphics[width=0.9\textwidth]{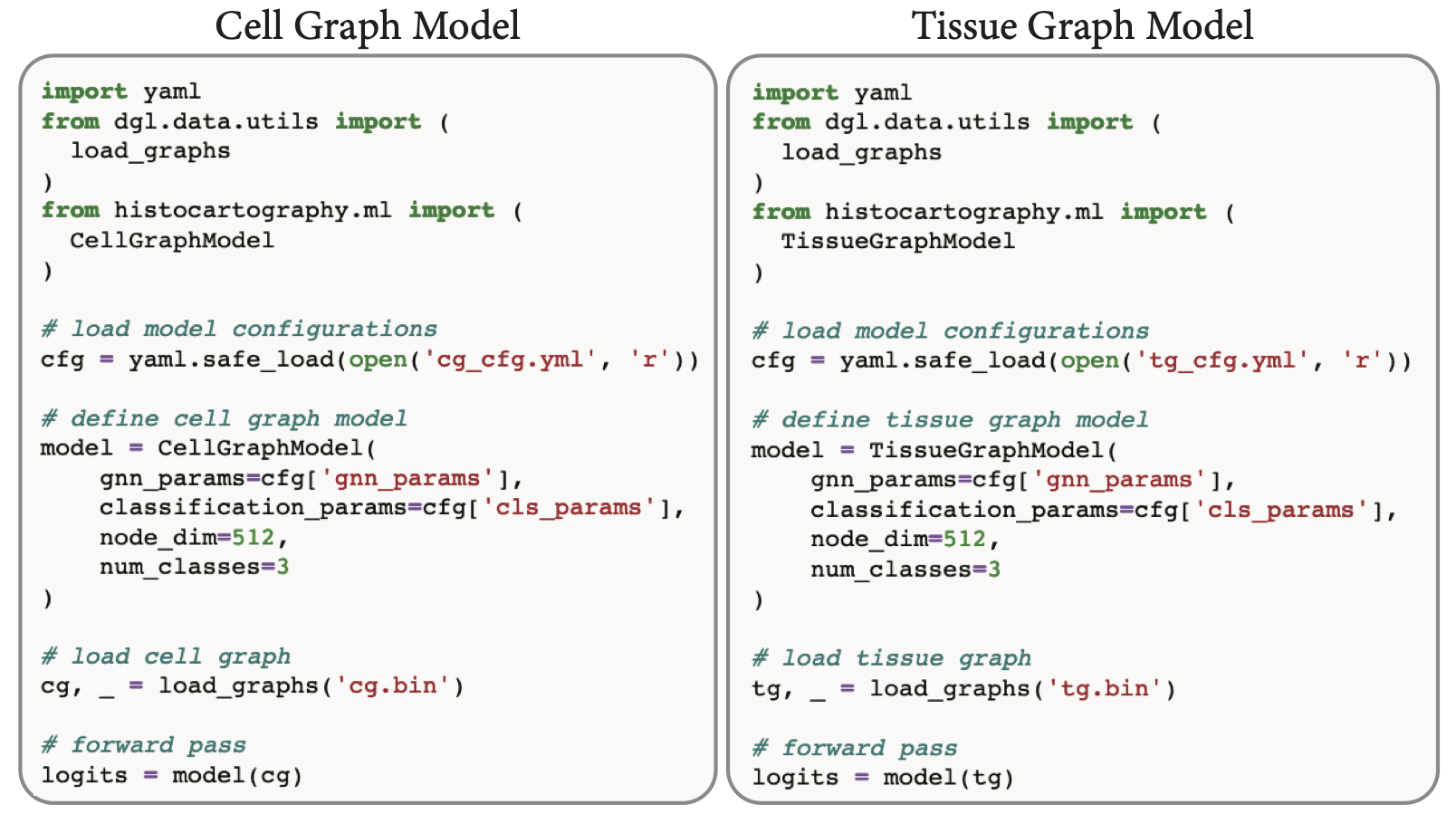}}
\vspace{-1em}
\caption{Implementation of the cell- (left) and tissue- graph (right) model by using the \gls{ml} modules in the $\histocartography$ \gls{api}}.
\label{fig:code3}
\vspace{-3em}
\end{figure}

\begin{figure}[h]
\centering
\centerline{\includegraphics[width=0.9\textwidth]{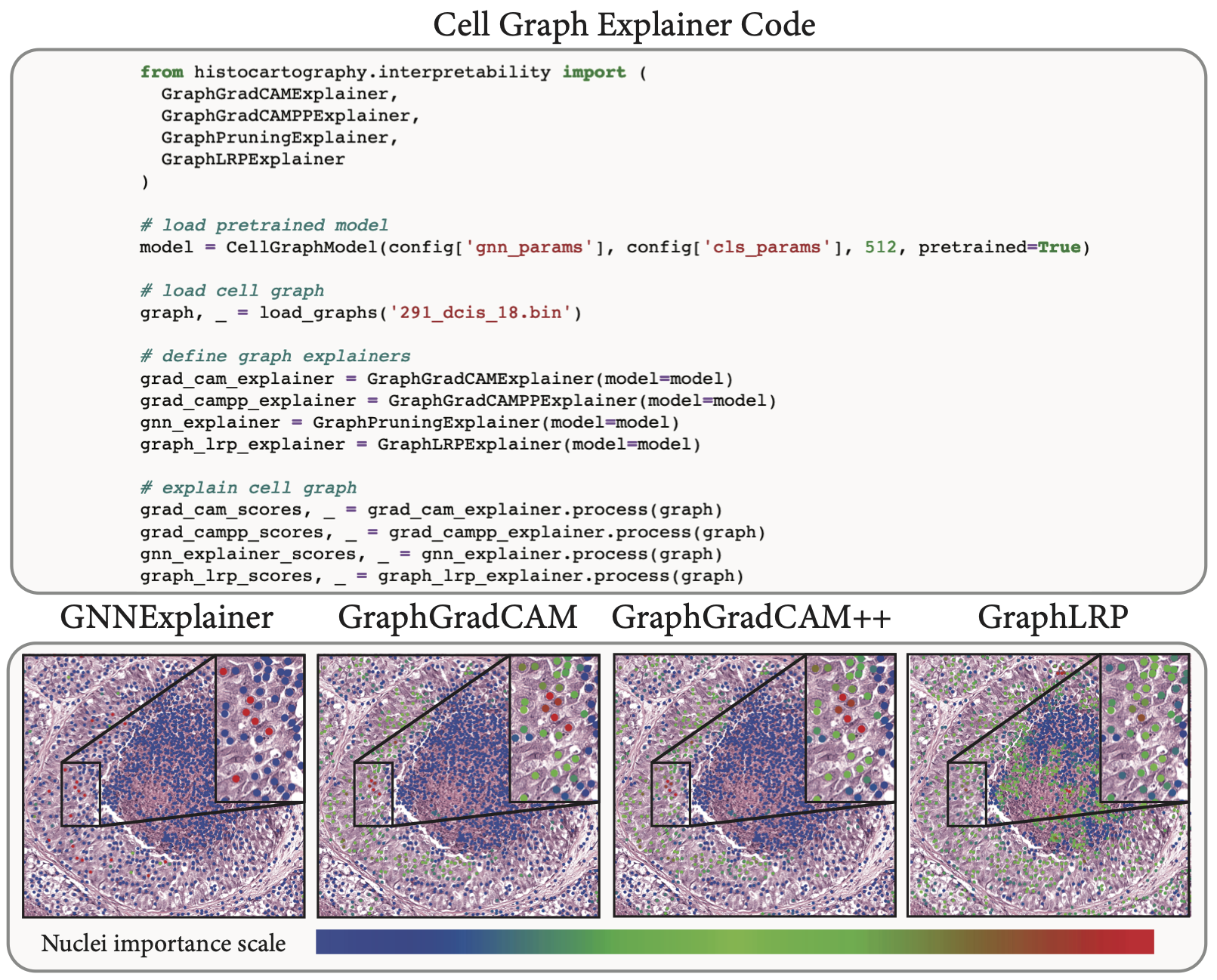}}
\vspace{-1em}
\caption{Implementation of graph explainers in $\histocartography$. The most important nodes are marked in red and the least important ones in blue.}
\label{fig:code4}
\end{figure}

In this section, we introduce the syntax to implement the functionalities of \textsc{HistoCartography}. Figure~\ref{fig:code1} presents code snippets to implement Vahadane stain normalization and tissue mask detection. Figure~\ref{fig:code2} shows the syntax for building cell- and tissue-graphs. Noticeably, these functionalities require only ten lines of code by using $\histocartography$, which could have otherwise required a few hundred lines. 
In Figure~\ref{fig:code3}, we present the syntax to declare and run a cell- and tissue-graph model. All the model parameters, \eg \gls{gnn} type, number of \gls{gnn} layers, can be adapted and fine-tuned using a configuration file. 
Finally, Figure~\ref{fig:code4} shows code snippets to use the graph explainability modules. All explainers follow a similar syntax with the same input and output types, making implementation and integration straightforward. 

\begin{table}[h]
\caption{Extended version of Table~\ref{tab:time}. Reported time to run $\histocartography$ core functionalities. CPU-only experiments were run on a single-core POWER8 processor, and GPU-compatible experiments were run on an NVIDIA P100 GPU. Time is reported in seconds.}
\label{tab:extended_time}
\centering
\resizebox{\columnwidth}{!}{%
\begin{tabular}{c|l|l|ccc|ccc}
  \toprule
  
  & & Modality & \multicolumn{3}{c|}{Tumor RoI} & \multicolumn{3}{c}{\gls{wsi}} \\
  \cmidrule{1-9}
  
  & & Size & $1000^2$ & $2500^2$ & $5000^2$ & $5000^2$ & $7500^2$ & $11000^2$ \\
  \cmidrule{1-9}

  \parbox[t]{2mm}{\multirow{11}{*}[-1.4ex]{\rotatebox[origin=c]{90}{Preprocessing}}} & \parbox[t]{2mm}{\multirow{4}{*}[-1.4ex]{\rotatebox[origin=c]{90}{Standard}}} & Vahadane Normalization & $1.77$ & $6.46$ & $29.03$ & $30.67$ & $68.27$ & $186.10$ \\ [0.1cm]

  & & Macenko Normalization & $0.80$ & $2.86$ & $11.19$ & $15.98$ & $32.37$ & $81.72$ \\ [0.1cm]

  & & Tissue Mast Detection & - & - & - & $1.04$ & $2.11$ & $8.09$ \\ [0.1cm]

  & & Feature Cube Extraction & $0.24$ & $1.61$ & $5.92$ & $6.27$ & $11.97$ & $29.79$ \\ [0.1cm]

  \cmidrule{2-9}

  & \parbox[t]{2mm}{\multirow{4}{*}[-1.4ex]{\rotatebox[origin=c]{90}{CG}}} & Nuclei Detection & $3.03$ & $12.93$ & $47.66$ & - & - & - \\ [0.1cm]

  & & Nuclei Concept Extraction & $2.95$ & $6.52$ & $27.94$ & - & - & - \\ [0.1cm]

  & & Deep Nuclei Feature Extraction & $0.10$ & $0.30$ & $1.28$ & - & - & - \\ [0.1cm]

  & & \gls{knn} Graph Building & $0.06$ & $0.20$ & $1.35$ & - & - & - \\ [0.1cm]

  \cmidrule{2-9}

  & \parbox[t]{2mm}{\multirow{3}{*}[-1.4ex]{\rotatebox[origin=c]{90}{TG}}} & Super-pixel Detection & $3.32$ & $17.84$ & $68.99$ & $31.50$ & $68.99$ & $183.54$ \\ [0.1cm]

  & & Deep Tissue Feature Extraction & $0.56$ & $2.99$ & $8.40$ & $4.17$ & $9.96$ & $20.54$ \\ [0.1cm]

  & & RAG Graph Building & $0.12$ & $2.04$ & $25.6$ & $6.33$ & $19.98$ & $85.73$ \\ [0.1cm]

  \cmidrule{1-9}
  
  \parbox[t]{2mm}{\multirow{3}{*}[-0.8ex]{\rotatebox[origin=c]{90}{ML}}} & & Cell-Graph Model & $0.028$ & $0.033$ & $0.040$ & - & - & - \\ [0.1cm]

  & & Tissue-Graph Model & $0.011$ & $0.015$ & $0.026$ & $0.039$ & $0.056$ & $0.069$ \\ [0.1cm]  

  & & HACT Model & $0.034$ & $0.041$ & $0.057$ & - & - & - \\ [0.1cm]
  \cmidrule{1-9}
  
  \parbox[t]{2mm}{\multirow{8}{*}[-0.8ex]{\rotatebox[origin=c]{90}{Explainers}}} & \parbox[t]{2mm}{\multirow{4}{*}[-1.4ex]{\rotatebox[origin=c]{90}{CG}}} & $\gnnexplainer$ & $12.00$ & $13.09$ & $35.33$ & - & - & - \\ [0.1cm]

  & & $\graphgradcam$ & $0.011$ & $0.022$ & $0.035$ & - & - & - \\ [0.1cm]

  & & $\graphgradcampp$ & $0.011$ & $0.023$ & $0.035$ & - & - & - \\ [0.1cm]

  & & $\graphlrp$ & $0.020$ & $0.024$ & $0.90$ & - & - & - \\ [0.1cm]

  \cmidrule{2-9}

  & \parbox[t]{2mm}{\multirow{4}{*}[-1.4ex]{\rotatebox[origin=c]{90}{TG}}} & $\gnnexplainer$ & $11.23$ & $11.28$ & $11.38$ & - & - & - \\ [0.1cm]

  & & $\graphgradcam$ & $0.011$ & $0.012$ & $0.018$ & $0.025$ & $0.030$ & $0.033$ \\ [0.1cm]

  & & $\graphgradcampp$ & $0.011$ & $0.013$ & $0.018$ & $0.026$ & $0.030$ & $0.033$ \\ [0.1cm]

  & & $\graphlrp$ & $0.011$ & $0.014$ & $0.016$ & $0.079$ & $0.085$ & $0.089$ \\ [0.1cm]

  \bottomrule
\end{tabular}
}
\end{table}

\subsection*{Handcrafted Feature Extraction}
In this section, we provide a comprehensive list of morphological and topological features which can be extracted per-entity by $\histocartography$. Morphological features include shape, size and texture properties, namely, entity area, convex area, eccentricity, equivalent diameter, euler number, length of the major and minor axis, orientation, perimeter, solidity, convex hull perimeter, roughness, shape factor, ellipticity, roudness. 
Texture properties are based on gray-level co-occurrence matrices (GLCM). Specifically, we extract the GLCM contrast, dissimilarity, homogeneity, energy, angular speed moment and dispersion. The topological features are based on the entity density computed as the mean and variance of entity crowdedness. 
These features can be computed for the most important set of entities highlighted by the graph explainability techniques, and utilized along with prior pathological knowledge to interpret the trained entity-graph models.

\end{document}